\documentclass[vanilla,keywords,pdf]{iucr}              % DO NOT DELETE THIS LINE
\usepackage{amsmath}
\usepackage{graphicx}
\usepackage{amssymb}

\begin{document}

\title{\emph{speckle-tracking}: a Software Suite for Ptychographic X-ray Speckle Tracking}
\shorttitle{speckle-tracking}
\maketitle

\cauthor[a,b]{Andrew J.}{Morgan}{morganaj@unimelb.edu.au (current affiliation: b)}{}
\author[c]{Kevin T.}{Murray}
\author[b]{Harry M.}{Quiney}
\author[c, d]{Sa\v{s}a}{Bajt}
\author[a,d,e]{Henry N.}{Chapman}

\aff[a]{Center for Free-Electron Laser Science, 
DESY, 
Notkestraße 85, 22607 Hamburg, Germany}
\aff[b]{ARC Centre of Excellence in Advanced Molecular Imaging, School of Physics, University of Melbourne, Parkville, Victoria 3010, Australia}

\aff[c]{DESY, Notkestrasse 85, 22607 Hamburg, Germany}

\aff[d]{Centre for Ultrafast Imaging, Universit{\"a}t Hamburg, Luruper Chaussee 149, 22761 Hamburg, Germany}

\aff[e]{Department of Physics, Universit{\"a}t Hamburg, 
Luruper Chaussee 149, 22761 Hamburg, Germany}

%\aff[f]{Faculty of Physics, University of Bialystok, K. Ciolkowskiego 1L, 15-245, Bialystok, Poland.}
%\aff[z]{?Please advise}

\shortauthor{Andrew J. Morgan \textit{et al.}}

%\keyword{x-ray diffraction}
%\keyword{diffuse scattering}
%\keyword{phase retrieval}
\keyword{software}
\keyword{wavefront metrology}
\keyword{speckle-tracking}
\keyword{ptychography}

\begin{abstract}
In recent years, x-ray speckle tracking techniques have emerged as viable tools for wavefront metrology and sample imaging applications. These methods are based on the measurement of near-field images. Thanks to the simple experimental set-up, high angular sensitivity and compatibility with low coherence sources these methods have been actively developed for use with synchrotron and laboratory light sources. Not only do speckle-tracking techniques give the potential for high resolution imaging, but they also provide rapid and robust characterisation of aberrations of x-ray optical elements, focal spot profiles and the sample position and transmission properties. In order to realise these capabilities, we require software implementations that are equally rapid and robust. To address this need, a software suite has been developed for the ``ptychographic x-ray speckle tracking technique'' -- an x-ray speckle based method suitable for highly divergent wavefields. The software suite is written in Python 3, with an OpenCL back end for GPU and multi-CPU core processing. It is accessible as a Python module, through the command line or through a graphical user interface and is available as source code under version 3 or later of the GNU General Public License.

\end{abstract}

\section{Introduction}

The X-ray Speckle-Tracking (XST) technique was introduced by \citeasnoun{Berujon2012a} and a little later by \citeasnoun{Morgan2012a}\footnote{No relation to the current author.} as a way to obtain a quantitative measurement of the phase of a wavefield, as induced by a transmitting object, for example. The general wavefront sensing principle of XST can be understood in terms of geometric optics, where it is assumed that light rays are directed normal to the isosurfaces of a wavefield's phase. 
An analyser object is placed in that wavefield and used to map the directions of ray paths. The analyser is typically chosen to be a thin sample with a random phase/absorption profile that produces intensity variations (or ``speckles'') some distance downstream, in the near field. These speckles are used as fiducial markers to determine the directions of the rays in the wavefield. 
%To determine the ray paths, a sample is placed in the path of the wavefield, typically a thin sample with a random phase/absorption profile, such that the local profile of the fine features in the wavefield, caused by transmission through the sample (which we term ``speckles''), acts as a fiducial marker for a ray's location. 
By measuring these speckle patterns at two detector distances, the lateral translation of speckles from one plane to the next allows the direction of a ray to be triangulated. This measurement strategy is termed the ``absolute configuration'' of XST, in the sense that the phase profile of the wavefield is recovered, regardless of whether those phase distortions originate from any particular upstream optics or sample. If, rather than two detector distances, two images are recorded at a single detector distance, one with and one without the presence of an additional sample, then the speckle translations encode the phase distortions caused by transmission through that sample rather than the wavefield itself. This measurement strategy is termed the ``differential configuration'' of XST, in the sense that the recovered phase profile is the difference between the phases of the wavefield before and after insertion of the sample. 

The ray path directions obtained through the ``absolution configuration'' XST method yield the two-dimensional (2D) phase gradient of the wavefield. 
Integrating these 2D phase gradients then provides the phase profile of the wavefield. Coupled with the wavefield's intensity profile, the wavefield can then be numerically propagated using wave optics from the measurement plane to any points of interest, such as the focal spot of a lens, the sample entrance surface or the lens pupil plane. 

Since 2012, a number of XST based methods have been developed, as reviewed by \citeasnoun{Zdora2018}. A more recent method is the Ptychographic X-ray Speckle Tracking (PXST) technique \cite{Morgan2019}. PXST was developed as a wavefront metrology technique that is optimised for the highly divergent wavefields produced by wedged multilayer Laue lenses \cite{Morgan2019a}. Note that PXST is not the only XST based technique that can be used in the presence of highly divergent wavefields, nor does it provide some of the sample imaging modalities that can be obtained with alternative approaches. 

In PXST, a dedicated speckle-producing analyser object is not used, and near-field speckle images are recorded at a fixed detector distance as a sample is scanned across the wavefield in the plane transverse to the optical axis. In the absolute configuration of Berujon's original proposal, the location of each ``speckle'' (here generated by the sample) in each image would be compared to those observed in a plane further upstream or downstream of the current detector plane. But this approach is impractical for highly divergent wavefields, since the footprint of the wavefield on the detector will be either much smaller or much greater than the detector's width. In the differential configuration of XST, the wavefield's ray paths could also have been determined by comparing the location of the observed speckles to those observed in an image of the same sample, but illuminated with an ideal wavefield, such that the ray path directions are known \textit{a priori}. But again, such an approach is impractical in cases where the optics to produce such an idealised wavefield are not available. In PXST, the missing reference image is instead recovered from the data itself, which is possible due to the high degree of redundancy in the data. This step -- extracting the virtual reference image from the data -- adds computational complexity to the wavefront reconstruction algorithm, but simplifies the experimental procedure. It also allows for large magnification factors in cases where it would otherwise be impractical to measure the reference image, which in turn allows for very precise measurements of a wavefront's phase and ray angles (to nano-radian accuracy). 

In PXST, the sample acts like the wavefront analyser but it need not necessarily have the properties that produce a speckle-like pattern in the near field. Indeed, as seen in the examples shown in Figs. \ref{fig:frameViewer} and \ref{fig:iters}, the approach generally works as long as features in the pattern (in that case, a Siemens star object) can be identified in several images. The result of the technique is a map of the phase and amplitude of the stationary wavefront (the ``illumination'') and a map of the near-field pattern of the sample as would be obtained in a perfect wavefield, with arbitrarily large field of view \cite{Morgan2019a}. As such, by separating the effects on the measurement of a stationary wavefield and a translated object, the method is similar to near-field ptychography \cite{Stockmar2013}\footnote{Note that for near-field ptychography, a diffuser is often employed to add structure to the illumination and improve robustness. Nevertheless, near-field ptychography is distinct from XST techniques.}. However, unlike that approach we rely upon ray optics rather than diffraction \cite{Morgan2019}. 

\begin{figure}
\includegraphics[width=8.88cm]{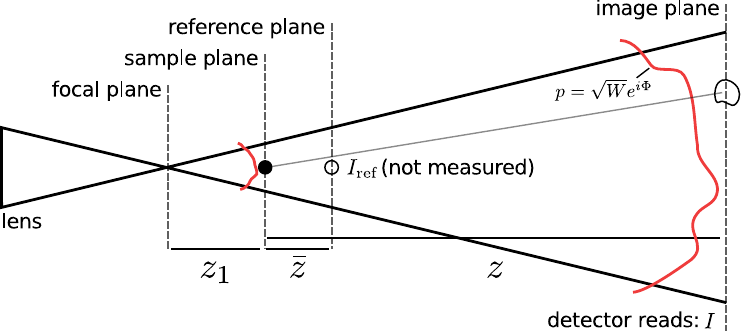}
%\vspace*{-1.0cm}
\caption{Schematic diagram for a projection imaging experiment. The 
illuminating beam propagates from left to right and the solid black lines 
indicate the boundaries of the illumination wavefront. The sample is depicted 
as a small black filled circle in the sample plane and as a black circle in the reference and image planes. The red lines depict the illumination's wavefront in the sample and image planes, which are not merely related by transverse magnification. The distorted shape of the circle in the image plane represents possible distortions of the speckle produced by the sample and the transverse phase gradients of the illumination.}
\label{fig:stem}
\end{figure}

In this article, we consider the general case of an imaging geometry corresponding to the projection imaging experiment as shown in Fig. \ref{fig:stem}.
Here, due to the beam diverging from the focus of a lens, the detector records a magnified near-field projection image (or hologram) of the sample. 
The equation relating the $n$\textsuperscript{th} recorded image ($I_n$) to the (un-recorded) virtual reference image ($I_\text{ref}$) in terms of the wavefront's phase profile ($\Phi$), is given by (Eq. 25 in \cite{Morgan2019}):
\begin{align}\label{eq:I}
I_n(\mathbf{x}) &\approx W(\mathbf{x}) I_\text{ref}(\mathbf{x} - \frac{\lambda z}{2\pi} \nabla \Phi(\mathbf{x}) - \Delta \mathbf{x}_n),
\end{align}
where $\mathbf{x}\equiv(x, y)$ is a vector in the plane transverse to the optical axis, $\nabla \equiv (\frac{\partial}{\partial x}, \frac{\partial}{\partial y})$ is the transverse gradient operator, the detector is a distance $z$ downstream of the sample, the wavelength of the monochromatic wavefield is $\lambda$, $\Phi$ and $W$ are the phase and intensity of the illumination's wavefront respectively in the plane of the detector and the sample displacement in the transverse plane corresponding to the $n$\textsuperscript{th} image is $\Delta \mathbf{x}_n$. 
The reference image corresponds to a magnified image of the sample one would have observed with plane wave illumination, covering the entire illuminated area of the sample throughout the scan, on a detector a distance $\bar{z}$ from the sample plane. In practice, for $W\approx 1$ and within the limits of the approximation in Eq. \ref{eq:I}, the reference image is simply a merge of each recorded image in the scan after correcting for the geometric distortions induced by the lens aberrations. The displacement of these overlaid images are given by the demagnified sample positions $\Delta \textbf{x}_n / M$, where $M$ is the magnification factor (see below). 

For an ideal lens system, producing an unaberrated diverging wavefield, $\bar{z} = \frac{z_1 z}{z_1 + z} = \frac{z}{M}$, where $z_1$ is the focus to sample distance and $M$ is the effective magnification factor. In Fig. \ref{fig:stem} we show the relative positions of the focal plane, the sample plane, the reference image plane and the imaging plane (where the detector is located). The central goal of the \textit{speckle-tracking} software suite is to solve the set of equations in Eq. \ref{eq:I} for $\Phi$ and $I_\text{ref}$ in terms of the recorded images $I_n$. 

Our intent is for the software described here to be accessible to imaging scientists and engineers, but who may not necessarily be experts in any particular field of imaging. In addition, we have also designed the software suite so that it might be integrated into the standard processing pipelines at beamlines and laboratory sources around the world. To these ends, we have developed three modes of user interaction with the basic routines of the suite: a python module interface, so that users have access to each of the low-level functions of the suite and may modify or replace such functions for different applications; a command-line interface, to enable automated processing of data-sets that may run in batch mode or remotely on a computing cluster; and a Graphical User Interface (GUI), designed for users who wish to interactively explore their data-sets and trouble-shoot for possible sources of error in the analysis. Python was chosen as the top level language due to its ease of use, its open source license and wide adoption by the scientific community. The more intensive calculations are executed in OpenCL, which can be run on multi-CPU hardware as well as Intel, NVIDIA and AMD based GPUs. OpenCL was chosen over alternatives such as CUDA, for example, which can only be executed on NVIDIA based GPUs. The OpenCL kernals are compiled and interfaced with the python routines at run-time via the PyOpenCL API \cite{kloeckner_pycuda_2012}. If the external dependencies of the suite are present, then the software does not require any installation or initial compilation. This further simplifies the set-up process and possible integration with other processing pipelines. 

Other groups have also distributed the source code for implementing a particular variant of the XST approach. For example, \citeasnoun{Paganin2018} have released the source code and experimental data for the ``single-image geometric-flow'' XST algorithm, available at https://github.com/labrieth/spytlab. \citeasnoun{Zdora} have also released the source code implementation for their Unified Modulated Pattern Analysis (UMPA), available at https://github.com/pierrethibault/UMPA.

\section{Main functions and overview}

The order in which the main functions, performed by the \textit{speckle-tracking} software suite, may be executed is illustrated in Fig. \ref{fig:flow}. Each of these functions can be called directly via the python module, where the input data and algorithm parameters are supplied by standard python objects. Corresponding to each of these functions is a command-line script of the same name. In each script, the input data is extracted from a CXI file (see section \ref{sec:cxi} below), with optional parameters read from a configuration file. The command-line script then executes the corresponding python function and returns the output into the same CXI file (unless otherwise configured). For each command-line script, there is a corresponding GUI, where the algorithm parameters can be set, the command executed and the output visually examined. The display routines are based on the \textit{pyqtgraph} project (www.pyqtgraph.org), which allows for fast and interactive visualisation of large data-sets. 

There are also a small set of additional GUIs (the first two are not shown in Fig. \ref{fig:flow}) that are not linked to a specific command-line script: 
\begin{itemize}
\item \textbf{hdf5\_viewer}, displays the contents of the CXI file and allows for basic visualisation of its constituent data-sets. 
\item \textbf{frame\_viewer}, for viewing individual images in correspondence with the transverse coordinates of the object. In this GUI one can manually exclude images from further analysis (see Fig. \ref{fig:frameViewer}).  
\item  \textbf{mask\_maker}, allows the user to manually edit the mask file of the detector. The mask is displayed as an overlay on top of the raw data, which allows the user to spot bad pixels that may not have been picked up from the automatic processing (e.g. via the \textbf{make\_mask} function). 
\end{itemize}

Equation \ref{eq:I} describes the relationship between the reference image, $I_\text{ref}$, and the $n$\textsuperscript{th} recorded image, $I_n$, as a geometric mapping defined in terms of the phase gradients $\nabla \Phi$ and the sample position $\Delta \textbf{x}_n$, so that Eq. \ref{eq:I} can be expressed compactly as $I_n(\textbf{x}) = W(\textbf{x}) I_\text{ref}(\textbf{u}(\mathbf{x}) - \Delta \textbf{x}_n)$, where $\textbf{u}$ is that component of the geometric mapping that remains constant from image to image, $\textbf{u}(\mathbf{x}) \equiv \mathbf{x} - \lambda z/ (2\pi) \nabla \Phi(\mathbf{x})$. In the \textit{speckle-tracking} suite, the reference image and the mapping function are solved for in an iterative update procedure that aims to minimise the target function:
\begin{align}\label{eq:error}
 \varepsilon(\mathbf{u}) &= \sum_n \int \frac{1}{\sigma^2_I(\mathbf{x})} \left[I_n(\mathbf{x}) - W(\textbf{x}) I_\text{ref}(\textbf{u}(\mathbf{x}) - \Delta \textbf{x}_n)\right]^2 d\mathbf{x},
\end{align}
where $\sigma^2_I(\mathbf{x})$ is the variance of the intensity signal at $\mathbf{x}$ over the $n$ recorded images.

\subsection{Initialisation}

\begin{figure}
\includegraphics[width=\textwidth]{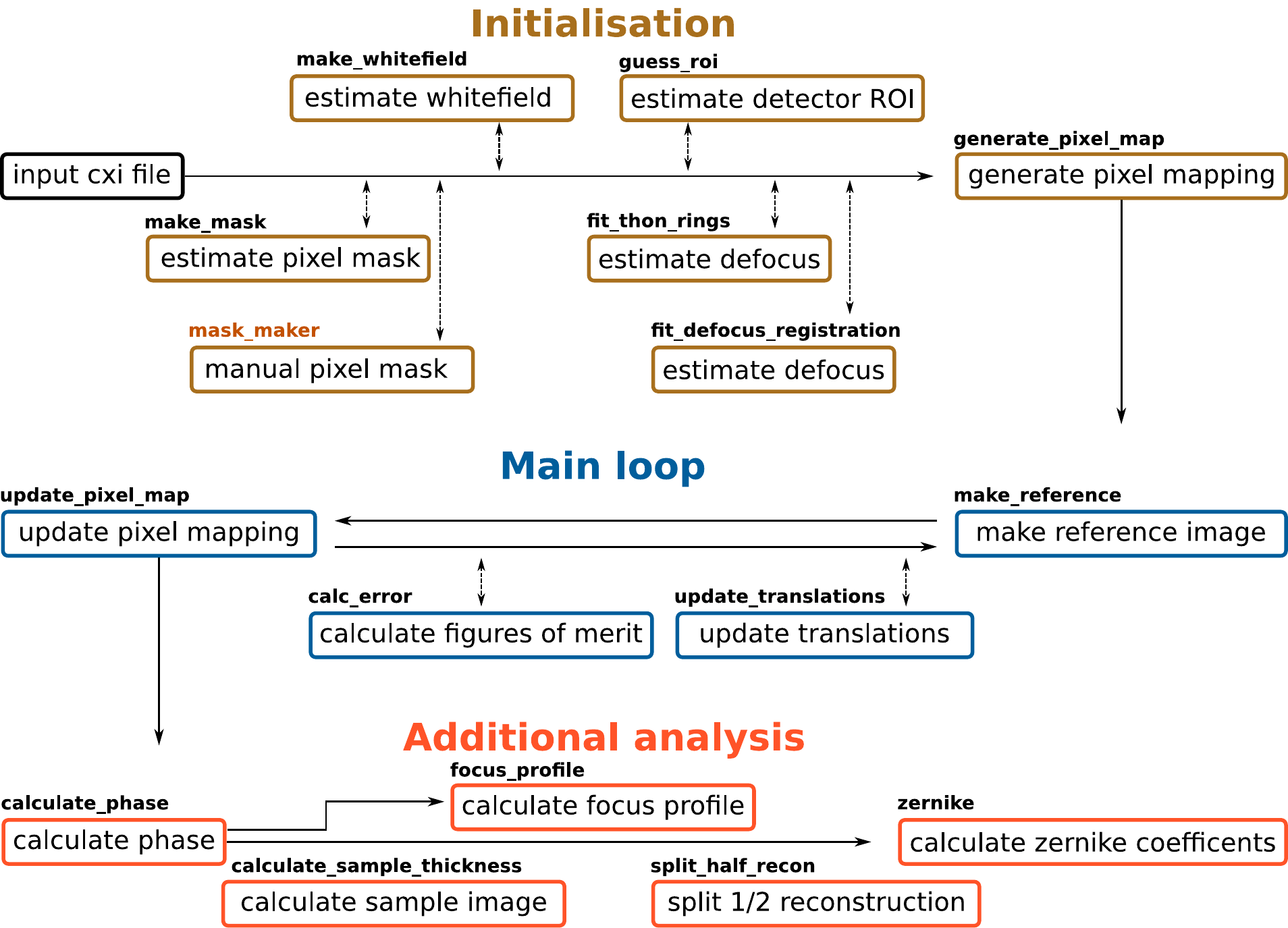}
%\includegraphics[width=8.88cm]{figures/cxi_file.pdf}
%\includegraphics[width=\textwidth]{figures/cxi_file.pdf}
%\vspace*{-1.0cm}
\caption{Flow diagram of the main operations in a PXST reconstruction. The name of the process for executing each step is displayed in bold above the description.}
\label{fig:flow}
\end{figure}

\begin{figure}
\includegraphics[width=\textwidth]{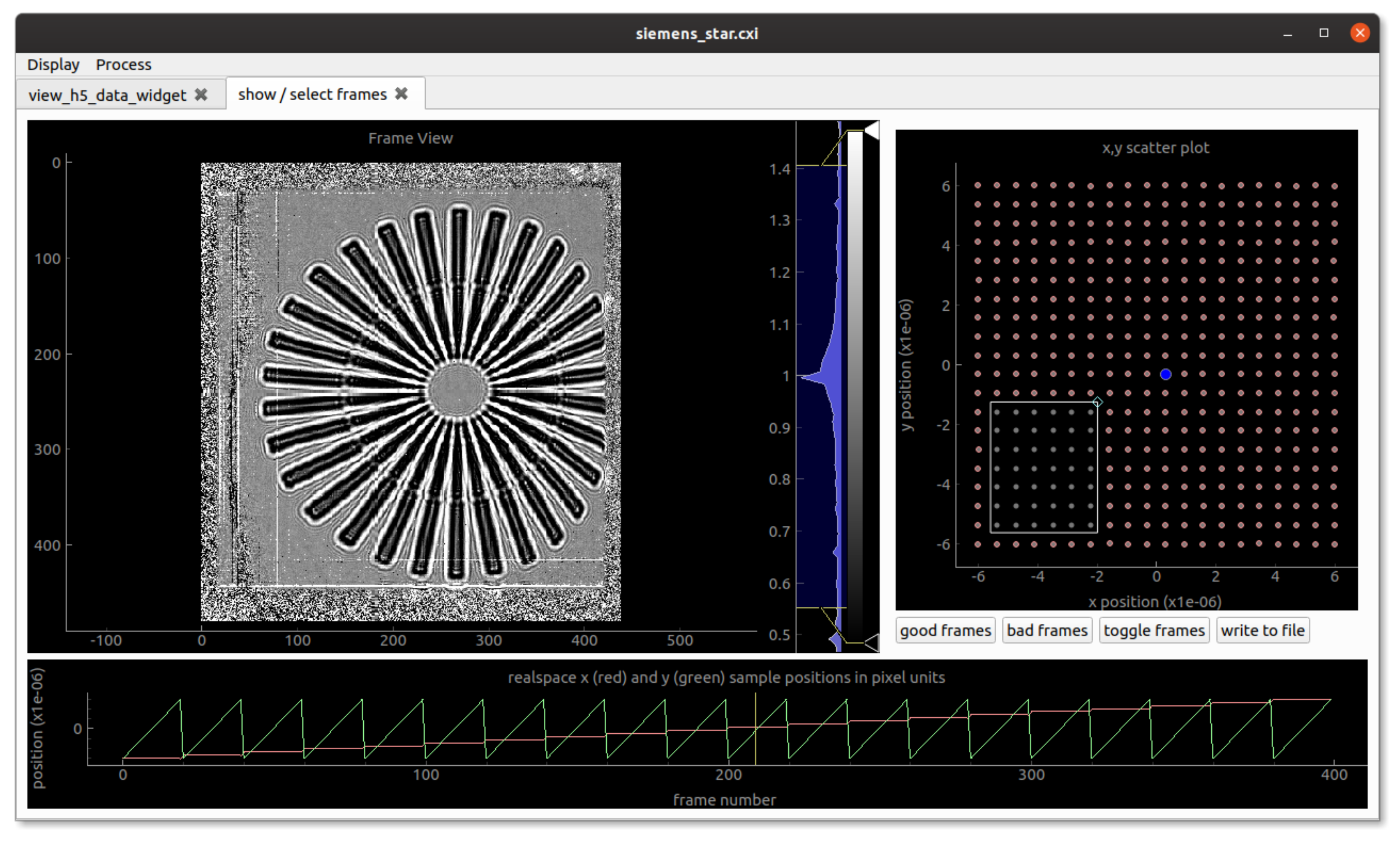}
%\includegraphics[width=8.88cm]{figures/cxi_file.pdf}
%\includegraphics[width=\textwidth]{figures/cxi_file.pdf}
%\vspace*{-1.0cm}
	\caption{Screen shot of the \textbf{frame\_viewer} GUI. The $x$ and $y$ coordinates of the sample translations ($\Delta \mathbf{x}_n$) are displayed both as line plots (bottom panel) and as a scatter plot (top right panel), the sample position corresponding to the current image is indicated by the vertical yellow line and the blue dot respectively. The current image is displayed in the top left panel, which will be divided by the white-field image, if present, in order to increase the contrast. One can scroll through the images by dragging the yellow line. Rather than read the entire data-set into system memory, each image is read from file in real-time so that very large data-sets can be accommodated. The white square in the top right panel can be used to select several frames to exclude from further analysis. The good (red dots) and bad (grey dots) status of a frame can also be toggled with the left mouse button.}
\label{fig:frameViewer}
\end{figure}

Ideally, one would already have several maps and quantities
%the pixel mask ($M$), whitefield image ($W$), sample defocus ($z_1$) and the detector's Region Of Interest (ROI) 
before performing the wavefront reconstruction. 
These are: the pixel mask ($M$), which indicates detector pixels common to all images $I_n(\mathbf{x})$ to exclude from subsequent analysis (equal to $1$ for good pixels and $0$ for bad); the white-field image ($W$) equal to the image that would be recorded without any sample present; the sample defocus ($z_1$); and the detector's region of interest (ROI), which is a rectangular region of the detector used in the analysis. In sitations where the ROI is non-rectangular, for example where the lens pupil is circular or elliptical, the pixels outside of this region can be excluded via the pixel mask.
In such a case, the only initialisation required in the processing pipeline is to generate the initial estimate for the mapping function $\mathbf{u}$, which depends on the focus to sample distance ($z_1$) and any estimated degree of astigmatism in the lens system (which can occur when separate MLLs focusing in each orthogonal direction are not adjusted into a confocal condition and which would give rise to different magnifications of the projection images in those directions). This step is performed by the \textbf{generate\_pixel\_map} function, as shown in the top right of Fig. \ref{fig:flow}. We refer to the discrete representation of the mapping function as the ``pixel mapping''. 
\onecolumn
\twocolumn

Naturally, each image is discretely sampled by the pixel array of the detector. For pixel $i$, $j$ of image $n$, the discrete representation of the image is given by $I[n, i, j] \equiv I_n(i \delta x, j \delta y)$, where $\delta x$ and $\delta y$ are the extent of a pixel along $x$ and $y$ respectively, and we have used square brackets to indicate discrete arguments; here, we ignore the fact that pixels do not provide point samples of the diffraction but rather integrate its intensity over the domain of the pixel response function. Because the reference image is not measured by the detector, the sampling of $I_\text{ref}$ is not constrained by the physical pixel geometry. Typically, we set the sampling period to equal the demagnified pixel size, so that $I_\text{ref}[i, j] \equiv I_\text{ref}(i \delta u, j \delta v)$, where $\delta u = \delta x / M$ and $\delta v = \delta y / M$. Likewise, the sample translations are converted to pixel units on the same grid, so that $\Delta i[n] = \Delta x_n / \delta u$ and $\Delta j[n] = \Delta y_n / \delta v$, where $\Delta \textbf{x}_n = (\Delta x_n, \Delta y_n)$. 
The function $\mathbf{u}(\mathbf{x})$ maps intensities from the detector grid to the reference image. Therefore, the sampling of the discrete representation of $\mathbf{u}$ is set by the detector geometry, and its values (coordinates in the reference image) are scaled by the chosen values of $\delta u$ and $\delta v$. 
This leads to the discrete representation of the mapping function: $u_x[i, j] \equiv \frac{1}{\delta u} u_x(i\delta x, j\delta y)$ and $u_y[i, j] \equiv \frac{1}{\delta v} u_y(i\delta x, j\delta y)$, where $\mathbf{u}(\mathbf{x}) \equiv (u_x(\mathbf{x}), u_y(\mathbf{x}))$ and we make a similar definition for the discrete mapping $\mathbf{u}[i, j] \equiv (u_x[i, j], u_y[i, j])$. Equation \ref{eq:I} can now be expressed in terms of these discrete quantities:
\begin{align}
 I[n, i, j] &= W[i, j] I_\text{ref}[u_x[i, j] - \Delta i[n], u_y[i, j] - \Delta j[n]].
\end{align}

As part of the software suite, there are a number of helper functions for estimating the data-sets necessary for initialising the main reconstruction loop. The function \textbf{make\_mask} attempts to locate and mask bad pixels automatically.
This is achieved by searching for pixels whose values deviate significantly from the median value of their neighbours during the scan. 
The output of this function can be viewed and manually modified by the \textbf{mask\_maker} GUI. The \textbf{make\_whitefield} function generates an estimate for the ``white-field'' image, by taking the median value of a pixel over the scan. The function \textbf{guess\_roi} will attempt to estimate the rectangular region of interest of the detector, by choosing the rectangular region containing most of the signal in the white-field image. This region should include what is sometimes referred as the ``brightfield'' or holographic region of the diffraction, where the white-field image differs appreciably from 0.

\onecolumn
\twocolumn
We provide two helper functions to estimate the focus to sample distance: \textbf{fit\_thon\_rings} and \textbf{fit\_defocus\_registration}. In \textbf{fit\_thon\_rings} the defocus and the degree of astigmatism are approximated by fitting a forward model to the Fourier power spectrum of the data. With sufficient source coherence and signal-to-noise, Fresnel fringes may be observed in each of the projection images. These fringes produce a ring like pattern in the Fourier power spectrum, the ``Thon'' rings \cite{thon1966defokussierungsabhangigkeit, spence2013high}; see for example Fig. 9 of \citeasnoun{Morgan2019a}. The spacing between these rings depends on the focus to sample distance along each of the transverse directions\footnote{see https://speckle-tracking.readthedocs.io/en/latest/thon\_rings.html for details.}.
This approach has the advantage that it is independent of the registered sample translations. 

In the other approach, \textbf{fit\_defocus\_registration}, a reference image of the sample is built for a range of magnification factors. The average magnification factors are given by $M_{x,y} = (z^{x,y}_1+z)/(z^{x, y}_1)$, where $z^{x, y}_1$ is the distance between the beam waist and the sample along the $x$ or $y$ axis respectively. If the estimated magnification factors are far from the true values, then each of the projection images will be misregistered when forming the reference image and the resulting image will lack contrast. The values for $z^x_1$ and $z^y_1$ that produce the greatest image contrast are then chosen. This approach will likely fail when the initial estimates for the sample translation vectors deviate significantly from the true values. The aim of these two functions is to produce an estimate for the defocus that is sufficiently accurate for the subsequent iterative refinement process to converge, after which, more accurate estimates for the defocus values can be obtained from the recovered phase profile.

\subsection{Main loop}
%First we form an initial estimate for the mapping function $\mathbf{u}$, which depends on the focus to sample distance ($z_1$) and the estimated degree of astigmatism in the lens system. This estimate is then used to generate an (imperfect) representation of the reference image ($I_\text{ref}$) by numerically inverting the effect of the mapping function on each image and merging the results. Features in the reference image are then matched with those in each of the recorded images and the value of the mapping function that most satisfies each 

Having formed an initial estimate for the pixel mapping function the iterative reconstruction cycle typically consist of; (i) generating the reference image; (ii) updating the pixel mapping between each image and the reference; (iii) updating the sample translation vectors; and (iv) calculating figures of merit. These four steps are described in the sections below. 
\onecolumn
\begin{figure}
\includegraphics[width=\textwidth]{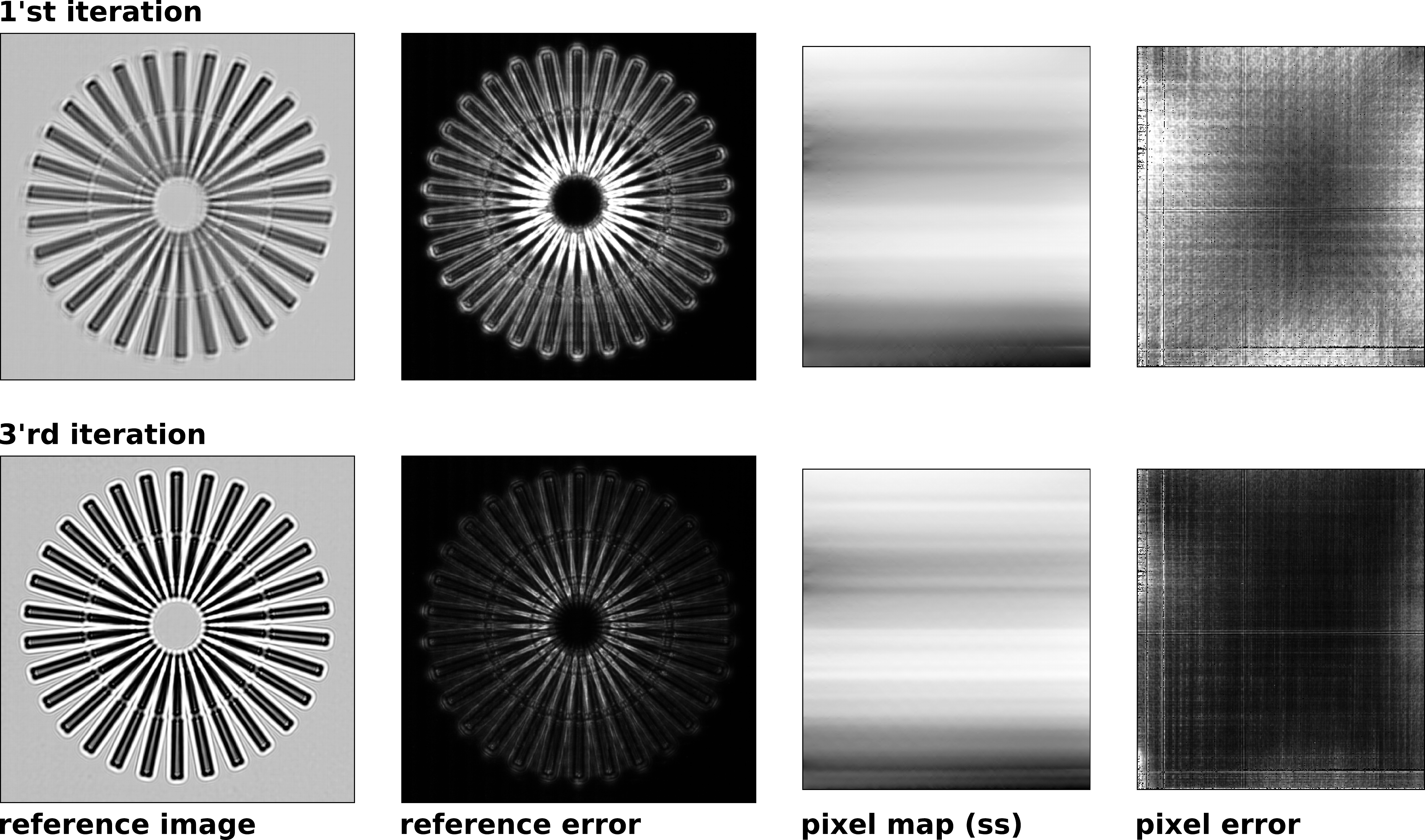}
%\includegraphics[width=8.88cm]{figures/cxi_file.pdf}
%\includegraphics[width=\textwidth]{figures/cxi_file.pdf}
%\vspace*{-1.0cm}
	\caption{The reference image (1'st column) and a component of the pixel map array (3'rd column), along the detector's slow scan axis (ss) (which corresponds to the vertical axis in this figure), for the 1'st and 3'rd iterations of the ``main loop''. In the 2'nd and 4'th columns we display the errors projected onto the reference image plane and the detector pixels, respectively. Each image is displayed on linear grey-scale colour map ranging from black (minimum value) to white (maximum value). The same colour map is used for images in the same column.}
\label{fig:iters}
\end{figure}
\twocolumn
This approach is an \textit{ad hoc} solution for minimising the error function (Eq. \ref{eq:error}). As such, convergence to the global solution is not guaranteed. Nevertheless, we have found that this simple approach performs well in many cases. For example, in Fig. \ref{fig:iters} we display the reference image and pixel map after the first and third iterations of the above procedure. This data was recorded at the HXN beamline of the NSLS-II (BNL, Upton, NY, US) with a Siemens star test sample as the wavefront analyser object, see \cite{Morgan2019a} for details. The error after the first loop was equal to $2.7\times 10^7$, after the third loop the error was reduced by a factor of 3 to $8.5\times 10^6$ and after 10 iterations the algorithm converged to an error of $5.9\times 10^6$. 

% error: 0 5.26e+07
% error: 1 2.88e+07
% error: 2 8.15e+06
% error: 3 6.85e+06
% error: 4 6.49e+06
% error: 5 6.24e+06
% error: 6 6.01e+06
% error: 7 5.94e+06
% error: 8 5.89e+06
% error: 9 5.87e+06

\subsubsection{Make reference image}
The equation for updating the reference image is described in Eq. 27 of \cite{Morgan2019}. We present the discrete representation of this equation as an operation in pseudo code:

\begin{align*}
&\text{initialise $I_\text{ref}$ and $w$ (an array of the same size) with zeros} \\
&\text{Loop over all images (index $n$)} \\
&\text{Loop over all pixels (indexes $i$ and $j$)} \\
 & \quad \text{add the mapped intensities, weighted by $W$, to $I_\text{ref}$:} \\
 & \quad I_\text{ref}[u[0, i, j] - \Delta i_n, u[1, i, j] - \Delta j_n]] \text{ += } M[i, j] W[i, j] I[n, i, j] \\
 & \quad \text{do the same for $W^2$, but add them to $w$:} \\
 & \quad w[u[0, i, j] - \Delta i_n, u[1, i, j] - \Delta j_n]] \text{ += } M[i, j] W^2[i, j] \\ 
 &\text{normalise the reference image by $w$ for all $[i,j]$:} \\
 & I_\text{ref} = I_\text{ref} / w
\end{align*}

where the symbols ``+='' signify the operation ``add the value on the right to the value on the left'' and $M$ is the pixel mask array which is $1$ for good pixels and $0$ for bad. In cases where the values of $u[0, i, j] - \Delta i_n$ and $u[1, i, j] - \Delta j_n$ are non-integer in the pseudo code above, we use a sub-pixel interpolation r\'{e}gime for assigning these values to $I_\text{ref}$ (see section \ref{sec:interp} for details). 

In the top left panel of Fig. \ref{fig:iters}, we show the reference image corresponding to the initial estimate for the pixel map as generated by \textbf{generate\_pixel\_map}. Looking closely, one can see that errors in the pixel mapping have resulted in placement errors, so that a given feature is misregistered when mapped to the reference image plane. After the third iteration, there are no visible placement errors and the Fresnel fringes near the edges of the Siemens star structure are now clearly visible. 

\subsubsection{Update pixel mapping}
The pixel mapping is updated by minimising the error function in Eq. \ref{eq:error} as a function of $\mathbf{u}$. This minimisation is performed in a brute-force search algorithm, where the error function is evaluated for every value of $\mathbf{u}$ within a predefined search window centred on the current estimate. We present the algorithm for updating the pixel mapping at pixel $[i, j]$, based on Eq. 28 in \cite{Morgan2019},  in pseudo code:
\begin{align*}
 & \text{initialise the error and variance (var) to zero} \\
 & \text{loop over search window along first axis (index $i'$)} \\
 & \text{loop over search window along second axis (index $j'$)} \\
 & \text{loop over all images (index $n$)} \\
 & \quad \text{add the error for image $n$ to error:} \\
 & \quad \text{error$[i', j']$ += } M[i,j] (I[n,i,j] - W[i, j] I_\text{ref}[i'-\Delta i_n, j'-\Delta j_n])^2 \\
 & \quad \text{calculate variance:} \\
 & \quad \text{var$[i', j']$ += } M[i,j] (I[n,i,j] - W[i, j])^2 \\
 & \text{normalise errors after all loops:} \\ 
 & \text{error} = \text{error} / \text{var} \\
 & \text{choose smallest normalised error:} \\
 & u[0, i, j] = i' \text{ and } u[1, i, j] = j' \text{ for $[i', j'] =$ argmin(error)} \\
\end{align*}
This update procedure is performed over the discrete pixel grid, i.e. for integer values of $i'$ and $j'$. To obtain sub-pixel resolution, one can evaluate the error for non-integer values of $i'$ and $j'$, as described in section \ref{sec:interp}, or set the ``quadratic\_refinement'' option to \texttt{True} in the \textbf{update\_pixel\_map} function. This quadratic refinement procedure was suggested by \citeasnoun{Zanette2014}, as part of the UMPA approach, and works by fitting a 2D paraboloid to the error profile in a $3\times 3$ pixel window about the minimum value of the error to obtain sub-pixel precision.

Since the mapping function is defined in terms of the gradient of a scalar function, $\mathbf{u} = \mathbf{x} - \lambda z/(2\pi) \nabla \Phi(\mathbf{x})$, the curl of the mapping function will be zero, $\nabla \times \mathbf{u}(\mathbf{x}) = 0$, which applies in cases where $\Phi$ is non-singular and continuous. In such cases the vector field $\mathbf{u}$ is said to be ``irrotational''. In the above procedure however, the vector components of $\mathbf{u}$, $u[0, i, j]$ and $u[1, i, j]$, were refined independently of each other. One can ensure that the mapping function is irrotational by first integrating $\mathbf{u}$, then by numerically evaluating the gradient to obtain an updated estimate for $\mathbf{u}$ with a curl of zero. The algorithm for integrating the pixel mapping is based on a least-squares conjugate gradient approach, also outlined in the paper by \citeasnoun{Zanette2014}, and can be enforced by setting ``integrate'' to \texttt{True} in the call to \textbf{update\_pixel\_map}.  

Finally, the pixel map can be filtered by setting the ``sigma'' parameter in the call to the update function. This step serves to regularise the pixel mapping, by filtering the pixel map with a Gaussian kernel, which can be useful in the early stages of the reconstruction to avoid artefacts that may arise from poor initial estimates of the reference image and the sample translations. 

In the third column of Fig. \ref{fig:iters}, we show the first component of the pixel map, $u[0, i, j]$. In the top row, the pixel mapping was generated using the above procedure, with the irrotational constraint applied and with a sigma value of 5 pixels, with the reference image shown in the first column. At the third iteration the sigma value was set to 1 pixel and thus finer details can be observed in the resulting map. Here, we have shown the pixel map after subtracting the linear term that arises from the estimated overall quadratic curvature of the phase. 

In our example of Fig. \ref{fig:iters}, a pair of MLLs was used to focus the beam \cite{Bajt2018}. Each MLL focuses the beam in one direction, like a cylindrical lens. We therefore expect that the dominant contribution to the lens aberrations would arise from placement and thickness errors in each layer of each of the 1D lenses. The orientation of each lens was
% For each of the 1D MLLs that make up the objective lens, we expect that the dominant contribution to the lens aberrations would arise from placement and thickness errors in each layer of the lens. Since each lens and the detector pixel axes are 
aligned along the pixel axes of the detector $[i,j]$, which also happen to be aligned to the $x$ and $y$ axes in the laboratory frame. Thus, assuming that each layer has a constant thickness in the direction perpendicular the focusing axis, we would expect each component of the pixel map would vary along one axis only. That is, $u_x(\mathbf{x})$ would vary along the $x$ axis and $u_y(\mathbf{x})$ would vary along the $y$ axis. One can see in the pixel maps of Fig. \ref{fig:iters}, that this is indeed the case for $u_x(\mathbf{x})$. Such a property would not be expected, for example, in an axially-symmetric lens, such as a compound refractive lens system.
 
\subsubsection{Update translations}
The procedure for updating the sample translation vectors follows the same logic as that for the pixel mapping update, as described in Eq. 29 of \cite{Morgan2019}. The error function of Eq. \ref{eq:error} can be evaluated as a function of $\Delta \mathbf{x}_n$ by performing the integral over $\mathbf{x}$ but not $n$. For a given image, the error is then evaluated for integer pixel shifts of that image within a predefined search window. Once again, sub-pixel precision is achieved by fitting a 2D paraboloid to the resulting error terms in $3\times 3$ pixel window about the minimum value. This process is then repeated for every image.  

\subsubsection{Calculate figures of merit}
The normalised error term in the integrand of Eq. \ref{eq:error} yields the error at a given pixel for a given image, $\varepsilon[n, i, j]$. The function \textbf{calc\_error} performs this calculation, then integrates the resulting errors: (i) over every image, yielding the ``pixel error'', $\varepsilon[i, j]$; (ii) over every pixel, yielding the ``frame error'', $\varepsilon[n]$; and (iii) over every pixel and image, yielding the ``total error'', $\varepsilon$. Following the same procedure used to form the reference image, one can also map the errors at each pixel for each image to the reference image plane, yielding the ``reference image error''. By projecting the error onto these different bases, it is easier to estimate the origin of potential artefacts in the reconstruction. 

For example, in the 2\textsuperscript{nd} and 4\textsuperscript{th} columns of Fig. \ref{fig:iters}, we show maps of the reference image error and the pixel error respectively. In the reference image error, we can see that errors are localised near the centre of the Siemens star projection image, where the Fresnel fringes from neighbouring edges begin to overlap. 

\subsection{Additional analysis}

The functions for generating the phase profile of the wavefront produced by the focusing optics (based on the recovered pixel map), subsequently for generating the profile of wavefront intensities near the focal plane and for calculating the aberration coefficients, are performed by the functions: \textbf{calculate\_phase}, \textbf{focus\_profile}, and \textbf{zernike}, respectively. 

The function \textbf{split\_half\_recon} takes every pixel of every image in the input data-set and randomly assigns it to one of two separate data-sets. Two pixel maps are then generated by comparing the existing reference image with each of halved data-sets. A histogram of the differences between these reconstructed pixel maps then provides an estimate for the underlying uncertainty in the original reconstruction. 

The function \textbf{calculate\_sample\_thickness} is an implementation of ``Paganin's algorithm'', where the thickness profile can be reconstructed from the reference image, see for example Eq. 61 in \cite{Paganin2019}, which is based on a low-order expansion the Transport of Intensity Equations (TIE) \cite{Teague1983}. This function is useful in that it provides a quick estimate for the sample's thickness profile (as opposed to the reference image which merely provides a projected in-line hologram of the sample) but is likely to be inaccurate for many applications. Other, more advanced, tools exist for this purpose; see for example, the X-TRACT software package by \citeasnoun{Gureyev2011}.

% flow diagram
% input.cxi --> estimate defocus --> generate pixel space coordinates --> [update reference image --> update pixel mapping -> regularise -> apply irrotational constraint --> update sample translations ] --> check figures of merit --> pixel mapping to phase --> generate propagation profile. 
%

% describe the work flow
% ?? key equations: main error function
% 
% sub-pixel interpolation, forward and inverse

\section{The Coherent X-ray Imaging (CXI) file format}\label{sec:cxi}

\begin{figure}
\includegraphics[width=12.cm]{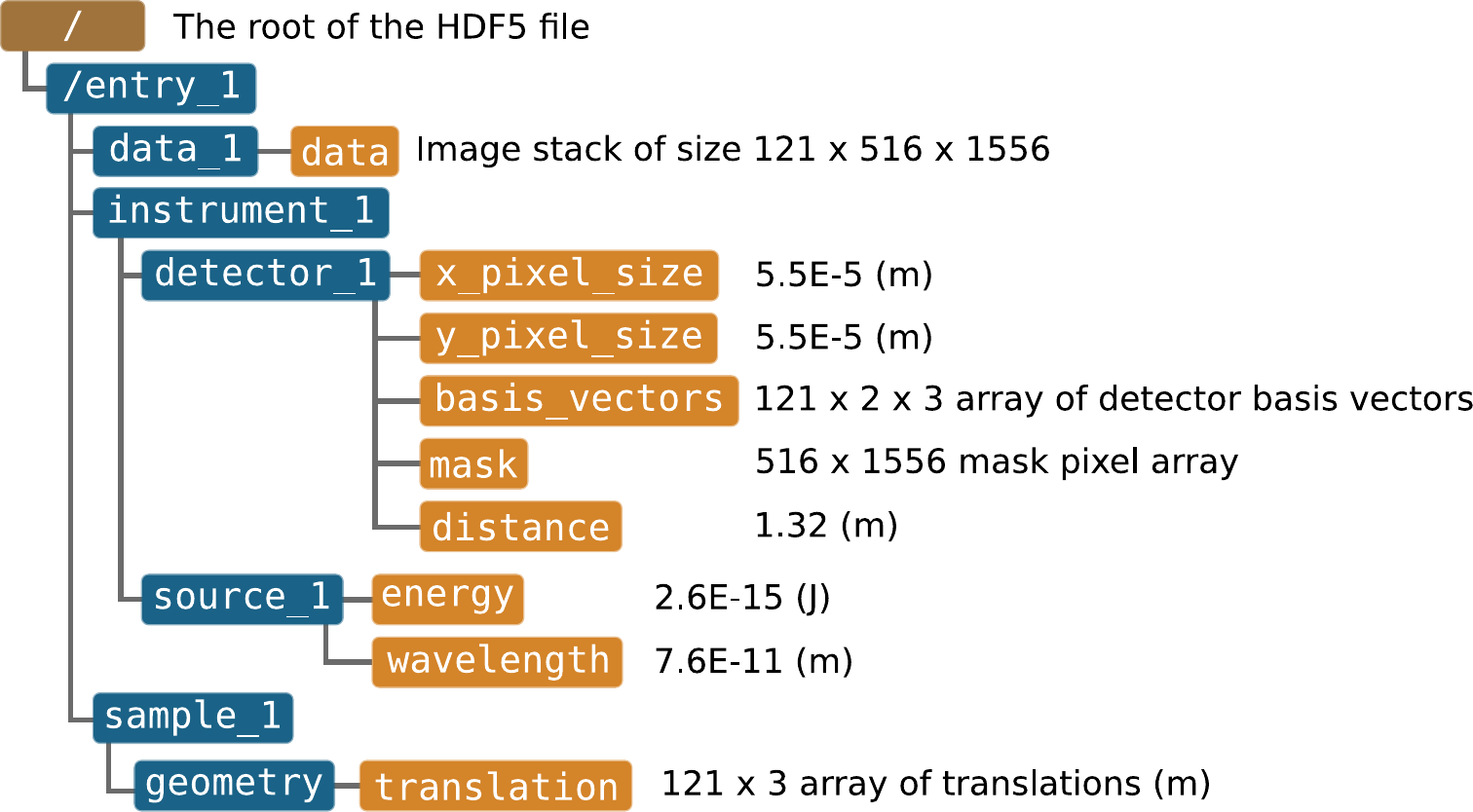}
%\includegraphics[width=8.88cm]{figures/cxi_file.pdf}
%\includegraphics[width=\textwidth]{figures/cxi_file.pdf}
%\vspace*{-1.0cm}
\caption{Diagram of a CXI file containing 121 images recorded on a 516 x 1556 pixel array detector. }
\label{fig:cxi}
\end{figure}

The \textit{speckle-tracking} software suite uses the CXI file format \cite{Maia2012} for input and output of key parameters and data-sets. The CXI format is based on the popular HDF5 format, which is a self-describing container for multidimensional data structures. The CXI format can be understood as simply a set of conventions for storing scientific data relating to coherent x-ray imaging in a HDF5 file. For example, the simplest input file for the \emph{speckle-tracking} program has the structure illustrated in Fig. \ref{fig:cxi}.
\onecolumn
\twocolumn
The CXI file format solves a number of common problems when translating data between programs and facilities: it establishes a default unit system (SI units), a default coordinate system, and a means of describing the relative orientation of the detector modules with respect to the laboratory frame of reference (via the \emph{basis\_vectors}). Providing a self-describing and unambiguous container for scientific data is also the aim of the NEXUS file format \cite{Maddison1997}, which is also built on top of the HDF5 format. However, perhaps due to its generality and scope, NEXUS-formatted files can be extremely complex, such that it can become difficult to extract meaningful data without specialised software. Thus, the CXI format was chosen as a suitable compromise between the aims of simplicity and an unambiguous representation of the data.

The output of the various programs within the \textit{speckle-tracking} suite can be stored into the same CXI file as the input, by default within the group called \textit{/speckle\_tracking}, or into another HDF5 file entirely, depending on the user configuration. Thus, the program's output data may, or may not, follow the default classes listed in the CXI standard. 

\subsection{Coordinate system}\label{sec:coord}

\begin{figure} 
\includegraphics[width=\textwidth]{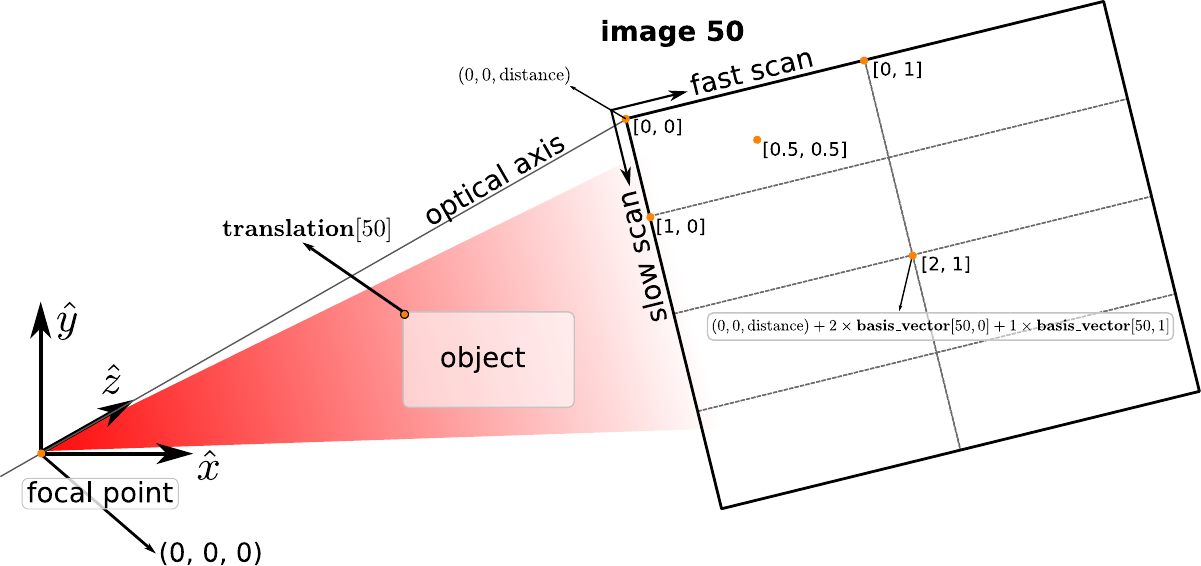}
%\vspace*{-1.0cm}
\caption{Diagram showing the relative positions of the focal point, the object and the detector pixels in terms of quantities stored in the CXI file. }
\label{fig:coord}
\end{figure}

\noindent One point of difference between the CXI standard and that adopted by the \textit{speckle-tracking} software, relates to the default origin of the coordinate system. In the CXI standard, the origin of the laboratory frame is centred on the sample of interest, with $\hat{z}$ parallel to the beam axis and $\hat{y}$ pointing against the direction of gravity in a right handed coordinate system. In many of our experiments, the sample's $z$ position is not known accurately in advance and often changes during the course of the experiment. Therefore, we have set the focal point of the optical system as the origin of the coordinate system, so that estimates for the sample's position may be refined \emph{post facto} without needing to offset the coordinates for the detector array and the focal point. Otherwise, we have followed the CXI standard as shown in Fig. \ref{fig:coord}. 

\onecolumn
\twocolumn

By default, the [0,0] pixel of the detector array is assumed to be centred on the optical axis with $z=\textit{distance}$ (located in the \textit{/entry\_1/instrument\_1/detector\_1} group of the CXI file). If this is not the case, then another data-set can be added to specify the corner position of the [0,0] pixel for each image. Currently, no part of the software suite actually makes use of such additional information, since the effect of an unknown translation (in the transverse plane) relative to the optical axis is to induce a corresponding offset in the reference image reconstruction in addition to a linear phase gradient on the wavefront reconstruction, which in any case is typically removed in the post reconstruction analysis. 

The $(x, y, z)$ coordinate of each pixel with respect to the pixel origin, i.e. pixel $[0,0]$ at $(0,0,distance)$ is defined by the \emph{basis\_vector} data-set. This is a three dimensional array, where the first dimension corresponds to the image number, the second dimension to the detector axes and the third to the three spatial dimensions. These vectors encode both the direction and magnitude of one step along the fast or slow scan axes of the detector array. Thus, the $(x, y, z)$ location of pixel $[2, 1]$ in image 50, can be obtained by starting at the location of pixel $[0,0]$ at $(0,0,distance)$, stepping twice along the slow scan axes of the detector, followed by one step along the fast scan axis $(0,0,distance) + 2 \times \mathbf{basis\_vector}[50, 0] + 1 \times \mathbf{basis\_vector}[50, 1]$.

\section{Sub-pixel interpolation}\label{sec:interp}
% show a figure depicting the bilinear interpolation routine

In the \textit{speckle-tracking} software suite, it is frequently the case that: (i) the value of an array must be evaluated at fractional pixel coordinates or (ii) that a value must be added to an array at a location with fractional pixel coordinates. After experimenting with a few common solutions to this problem, we have elected to use the bilinear interpolation formulation. This is a simple extension of 1D linear interpolation to the problem of interpolating a function of two variables.

If the values of a 2D array represent discrete samples of a continuous function over a Cartesian grid, with a fixed and equal pixel spacing along both dimensions, then bilinear interpolation reduces to a simple form. Consider the four values of some array, $f[0, 0]$, $f[1,0]$, $f[0,1]$ and $f[1,1]$, which are point samples of the continuous function $f(x, y)$ at each corner of a unit square. The value of $f$ at some point $[x, y]$ that is within this unit square can be approximated by:
\begin{align}
 f(x, y) &\approx 
 \begin{bmatrix}
 1 - x & x 
 \end{bmatrix} 
 \cdot 
 \begin{bmatrix}
 f[0, 0] & f[0, 1] \\
 f[1, 0] & f[1, 1]
 \end{bmatrix}
 \cdot
 \begin{bmatrix}
 1 - y \\
 y 
 \end{bmatrix}. 
\end{align}

\begin{figure} 
\includegraphics[width=\textwidth]{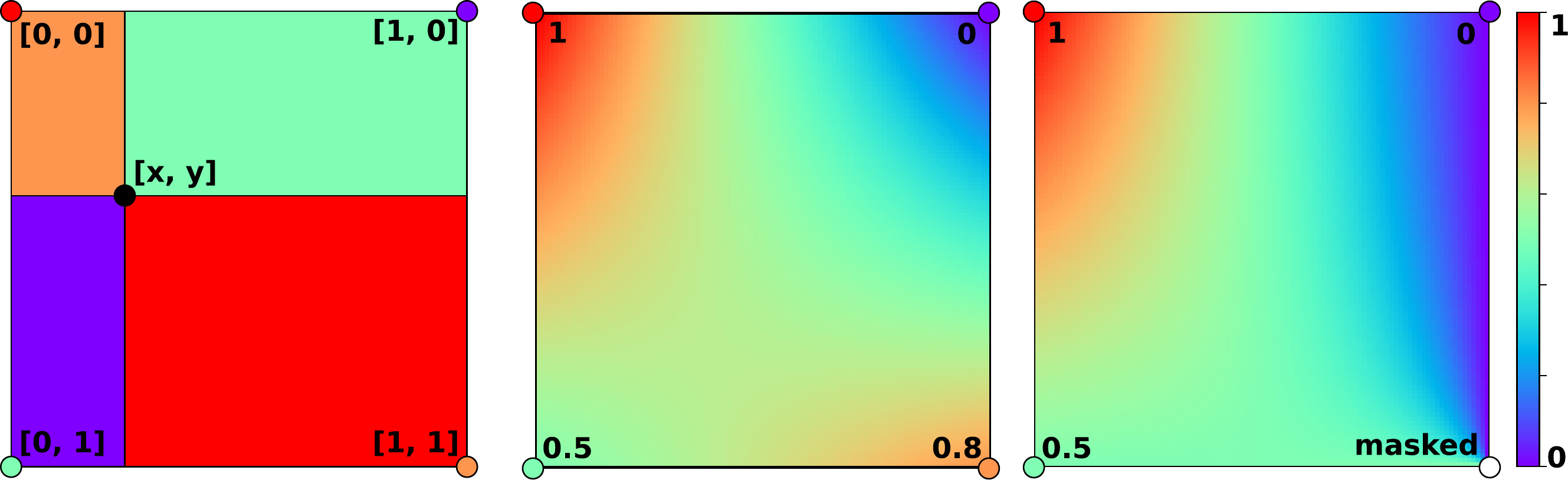}
%\vspace*{-1.0cm}
\caption{\textbf{Left:} Diagram showing the relative weights corresponding to the values at each corner of the unit-square. The weights corresponding to each point are equal to the area occupied by the rectangle diagonally opposite and are colour coded to illustrate this. \textbf{Middle:} The interpolated values for all points within the unit-square, given the values shown at each vertex. \textbf{Right:} Same as middle, but for the case where the value at the bottom right corner is unknown.}
\label{fig:interp}
\end{figure}

This corresponds to subdividing the unit square into four rectangular segments, centred on the point $[x, y]$, so that the weighting given to a particular corner value corresponds to the area occupied by the rectangle diagonally opposite its position, as shown in the left panel of Fig. \ref{fig:interp}. Since the total area of the four rectangles is equal to that of the square, the four weighting factors sum to 1. However, if one of the corner values is unknown, then this value can be excluded from the interpolation routine and the remaining values can be renormalised. Setting $f_M \equiv f\times M$, were $M$ is a mask array consisting of ones (for good values of $f$) and zeros (for bad or missing values of $f$), then the values of $f(x, y)$ are approximated by:
\onecolumn
\twocolumn

\begin{align*}
 f(x, y) \approx 
 &\begin{bmatrix}
 1 - x & x 
 \end{bmatrix} 
 \cdot 
 \begin{bmatrix}
 f_M[0, 0] & f_M[0, 1] \\
 f_M[1, 0] & f_M[1, 1]
 \end{bmatrix}
 \cdot
 \begin{bmatrix}
 1 - y \\
 y 
 \end{bmatrix} \\
 / 
 &\begin{bmatrix}
 1 - x & x 
 \end{bmatrix} 
 \cdot 
 \begin{bmatrix}
 M[0, 0] & M[0, 1] \\
 M[1, 0] & M[1, 1]
 \end{bmatrix}
 \cdot
 \begin{bmatrix}
 1 - y \\
 y 
 \end{bmatrix}.
\end{align*}
In Fig. \ref{fig:interp} (right panel), we show an example of the distribution of values when the bottom right corner value is unknown, for comparison with the case in the central panel where all four corner values of $f(x, y)$ are known. 

Conversely, when assigning a given value of $f$ at the fractional coordinate $[x, y]$ to the discrete array at each of the corner positions, the same weighting scheme is employed. If there exists more than one value of $f(x, y)$ to interpolate onto the discrete grid, then the weighting factors are stored in an array for renormalisation after all points have been assigned. 

One disadvantage of this approach is that it is not invertable. For example, interpolating an array of values onto a regular grid that is translated with respect to the original grid by some fraction of a pixel, followed by a corresponding shift in the opposite direction, does not preserve the original values of the array. Furthermore, interpolated values can have a discontinuous gradient when crossing the boundary between pixels, even when the underlying function $f(x, y)$ has continuous first derivatives. 

At the same time, however, this approach has a number of advantages. For example, sharp edges or points of high contrast in the original array do not produce artefacts that span the entire domain of the interpolated array, which can arise when using Fourier based interpolation routines. Because the interpolation is local, only pixel values in the immediate neighbourhood of the point $[x, y]$ are needed to compute the interpolated value. This property is particularly advantageous for parallel computations. In addition, masked values can be easily accommodated by renormalising each of the weighting terms. 
%\section{Utilities}

\section{Software availability}

In order to maximise the utility and accessibility of the software suite, \textit{speckle-tracking} is available under version 3 or later of the GNU general public license. This allows for other software projects to incorporate and modify all or part of this program into their own processing pipeline. The software can be downloaded at https://github.com/andyofmelbourne/speckle-tracking.

\section{Documentation}

High-level documentation for the project, including tutorials and installation instructions, can be found at https://speckle-tracking.readthedocs.io. For help with the command-line programs, simply pass \texttt{--help} as an argument when calling the program. This will provide a brief description of the program and the list of parameters accepted in the configuration file. Similar descriptions are provided in the ``doc string'' of the python functions. In each of the GUIs, descriptions of the input parameters can by viewed in a pop-up window, which is displayed when hovering the mouse over the parameter name.

\section{Available speckle tracking data-sets}

Currently, there are three experimental data-sets available to download on the CXIDB (https://www.cxidb.org), data-sets 134-136. Each data-set is stored in the CXI file format, which can be used immediately as input for the \textit{speckle-tracking} software suite. For testing with other programs, individual data-sets and parameters within the file can be extracted and converted to other formats using one of the many HDF5 APIs (available at https://www.hdfgroup.org). Tutorials\footnote{Found here: https://speckle-tracking.readthedocs.io/en/latest/\#tutorials} have been written for two of the available data-sets, which is the recommended way to reproduce the results in this article and to become familiar with the software suite.

% All of the command line programs in the \textit{speckle\_tracking} software suite accept as input a configuration file and CXI file. 
% solves confusion with coordinate systems and units. 
% most commands accept an ini file and a cxi file as input. 

% file format
% based on h5 --> popular self describing container format: nexus
% coordinate system
% example file 

% main routines
% interpolation
\section{Acknowledgements}

Funding for this project was provided by the Australian Research Council Centre of Excellence in Advanced Molecular Imaging (AMI), the Gottfried Wilhelm Leibniz Program of the DFG and the Cluster of Excellence 'CUI: Advanced Imaging of Matter' of the Deutsche Forschungsgemeinschaft (DFG) - EXC 2056 - project ID 390715994. This research used the HXN beamline of the National Synchrotron Light Source II, a U.S. Department of Energy (DOE) Office of Science User Facility operated for the DOE Office of Science by Brookhaven National Laboratory under Contract No. DE-SC0012704.

\bibliographystyle{iucr}
\bibliography{speckle_code}

\end{document}